\documentclass[traditabstract]{aa}
\usepackage{epsfig}    
\usepackage{lscape}
\usepackage{natbib}
\bibpunct{(}{)}{;}{a}{}{,}    
\usepackage{graphics}
\usepackage{graphicx,graphics}
\usepackage{color} 
\usepackage{times}
\usepackage{amsmath}
\usepackage{amssymb}
\begin{document}

\title {Two types of distribution of the gas velocity dispersion of MaNGA galaxies}

\author {
         L.~S.~Pilyugin\inst{\ref{MAO}}   \and 
         I.~A.~Zinchenko\inst{\ref{LMU},\ref{MAO}} \and
         M.~A.~Lara-L\'{o}pez\inst{\ref{AOP}} \and
         Y.~A.~Nefedyev\inst{\ref{KGU}} \and
         J.~M.~V\'{i}lchez\inst{\ref{IAA}}
         }
\institute{Main Astronomical Observatory, National Academy of Sciences of Ukraine, 27 Akademika Zabolotnoho St, 03680, Kiev, Ukraine \label{MAO}  \and
Faculty of Physics, Ludwig-Maximilians-Universität, Scheinerstr. 1, 81679 Munich, Germany \label{LMU} \and
Armagh Observatory and Planetarium, College Hill, Armagh BT61 9DG, Northern Ireland, UK \label{AOP} \and 
Kazan Federal University, 18 Kremlyovskaya St., 420008, Kazan, Russian Federation \label{KGU} \and 
Instituto de Astrof\'{\i}sica de Andaluc\'{\i}a, CSIC, Apdo 3004, 18080 Granada, Spain \label{IAA} 
}

\abstract{
  The distribution of the gas velocity dispersion $\sigma$ across the images of 1146 MaNGA galaxies
  is analyzed. We find that there are two types of distribution of the gas velocity dispersion
  across the images of galaxies: $(i)$ the distributions of 909 galaxies show a radial symmetry with
  or without the $\sigma$ enhancement  at the center ("R distribution,"
  radial symmetry in the $\sigma$ distribution) and $(ii)$ distributions with a band of enhanced
  $\sigma$ along the minor axis in the images of 159 galaxies with or without the $\sigma$ enhancement
  at the center ("B distribution,"  band in the $\sigma$ distribution).
  The $\sigma$ distribution across the images of 78 galaxies cannot be reliable classified.
  We select 806 galaxies with the best defined characteristics (this sample includes 687 galaxies with
  R distribution and 119 galaxies with B distribution)  and compare the properties of galaxies with
  R and B distributions.
  We find that the median value of the gas velocity dispersion $\sigma_{m}$ in galaxies with B distribution
  is higher by around 5 km/s, on average, than that of galaxies with R distribution. 
  The optical radius $R_{25}$ of galaxies with B distribution is lower by around 0.1 dex, on average,
  than that of galaxies with similar masses with R distribution. Thus the properties of
  a galaxy are related to the type of  distribution of the gas velocity dispersion $\sigma$ across its image. 
  This suggests that the presence of the band of the enhanced gas velocity dispersion can be an indicator of
  a specific evolution (or a specific stage in the evolution) of a galaxy.   
}

\keywords{galaxies: ISM -- galaxies: kinematics and dynamics -- galaxies: structure}

\titlerunning{Gas velocity dispersion in galaxies}
\authorrunning{Pilyugin et al.}
\maketitle

\section{Introduction}

In a previous study, we considered the circumnuclear regions in 161 star-forming (SF) galaxies from the Mapping Nearby Galaxies at Apache
Observatory (MaNGA)  survey \citep{Bundy2015,Albareti2017} with four radiation distribution configurations \citep{Pilyugin2020b}.
The spaxel spectra were classified as  active galactic nucleus-like (AGN-like),  H\,{\sc ii}-region-like (or SF-like), and   
intermediate (INT) spectra according to their positions in the Baldwin-Phillips-Terlevich \citep[BPT, ][]{Baldwin1981} diagram. 
We found that the AGN-like and INT radiation in the circumnuclear region is accompanied
by an enhancement in the gas velocity dispersion $\sigma$, in the sense that the  gas velocity dispersion in
a galaxy decreases with galactocentric distance up to some radius and remains approximately constant beyond this radius.
The radius of the area of the AGN-like and INT radiation (radius of influence of the AGN on the radiation) 
is similar to the radius of the area with enhanced gas velocity dispersion, and the central gas velocity dispersion
$\sigma_{c}$ correlates with the luminosity of the AGN+INT area. 
This suggests that the enhancement of the gas velocity dispersion at the center of a galaxy can be attributed to the AGN activity.
 
A correlation between the gas velocity dispersion and the diagnostic emission line ratios of the BPT diagram is revealed in the ultraluminous and
luminous infrared galaxies in the absence of any contribution from an AGN. This correlation is interpreted
as a signature of shock excitation, where shocks are driven by galaxy mergers or interactions
\citep{Monreal2006,Monreal2010,Rich2014,Rich2015}.

Here we construct the maps of the gas velocity dispersion for a large sample of  late-type MaNGA galaxies
with the aim of examining the properties of the gas velocity dispersion fields beyond the circumnuclear regions.
The parameters of emission lines in each spaxel spectrum are estimated using a Gaussian fit to the line profiles.
The sigma of the best-fit Gaussian  of the  H$\alpha$ line  is converted into the gas velocity dispersion.

This paper is organized as follows. The data are described in Section 2. In Section 3, the distributions of the
gas velocity dispersion  of MaNGA galaxies are discussed, and Section 4 provides a brief summary.

\section{Data}

Our current study is based on  data obtained from the public spectroscopic observations  from 
the Sloan Digital Sky Survey Data Release 15 (SDSS DR15) MaNGA  survey \citep{Bundy2015,Albareti2017} and described in \citet{Pilyugin2020b}.
In brief, the spectrum of each spaxel was reduced as indicated in \citet{Zinchenko2016} and \citet{Sakhibov2018}.
The stellar background was fitted using the public version of the STARLIGHT code
\citep{CidFernandes2005,Mateus2006,Asari2007}, which was adapted for execution in
the NorduGrid Advanced Resource Connector (ARC)\footnote{http://www.nordugrid.org/} environment of the Ukrainian National Grid.  
The library of synthetic simple stellar population (SSP) spectra from the evolutionary synthesis
models in \citet{Bruzual2003} were used. The reddening law from \citet{Cardelli1989} with $R_V = 3.1$ was adopted.
The obtained stellar radiation contribution was subtracted from the observed spectrum. 

The  emission lines in each spaxel spectrum were estimated using a Gaussian fit to the line profiles. The emission line parameters 
include the central wavelength $\lambda_{0}$, the sigma $\sigma$, and the flux $F$. For each spectrum, we measured the 
[O\,{\sc ii}]$\lambda\lambda$3727,3729, H$\beta$, [O\,{\sc iii}]$\lambda$5007, H$\alpha$, and [N\,{\sc ii}]$\lambda$6584 emission lines. 
Only those spaxel spectra where all the used lines were measured with a signal-to-noise ratio S/N $> 3$ were taken into consideration.  
The central wavelength of the H$\alpha$ line was converted into the line-of-sight velocity, and  
the sigma of the best-fit Gaussian $\sigma_{{\rm H}\alpha}$ was converted into the observed gas velocity dispersion $\sigma_{obs}$.
Since the values of the observed (non-corrected-for instrumental profile) gas velocity dispersion are used throughout the paper, we will use the notation $\sigma$ instead of  $\sigma_{obs}$. The gas velocity dispersion in the galaxy will be specified by
the median value of the gas velocity dispersion $\sigma_{m}$ of the individual spaxels.
  
The measured emission line fluxes were corrected for interstellar reddening using the theoretical H$\alpha$/H$\beta$ ratio and the reddening
function from \citet{Cardelli1989} for $R_{V}$ = 3.1.  We assumed $C_{{\rm H}{\beta}} = 0.47A_{V}$ \citep{Lee2005}.
The excitation properties of each spaxel spectrum (main excitation mechanism: starburst or AGN)
were determined using the diagnostic line ratio diagram proposed by \citet{Baldwin1981} (BPT classification diagram).
The spectra located to the left (below) the demarcation line from \citet{Kauffmann2003} are referred to as the SF-like or  H\,{\sc ii} region-like spectra, 
those located to the right (above) the demarcation line from \citet{Kewley2001} are referred to as the AGN-like spectra, and  
the spectra located between those demarcation lines are  the intermediate (INT) spectra.
The configuration of the radiation distribution in the circumnuclear region of each galaxy was classified according to \citet{Pilyugin2020b}:
(i) the circumnuclear region of the  SF or H\,{\sc ii} type where the central area involves the spaxels with the  H\,{\sc ii}-region-like radiation only;
(ii) the circumnuclear region of the  INT type if the radiation at the center of the galaxy is the intermediate type;  
(iii) and the circumnuclear region of the AGN type where the innermost region of the AGN-like radiation is surrounded by a ring of radiation
of the intermediate type. The circumnuclear regions of the AGN type are divided into two subtypes, the LINER circumnuclear regions 
and the Seyfert circumnuclear regions, using the dividing line between Seyfert galaxies and LINERs defined by \citet{CidFernandes2010}.

The geometrical parameters of the galaxy were determined from the surface brightness distribution
in the same way as in our previous studies  \citep{Pilyugin2014,Pilyugin2017,Pilyugin2018}. 
The surface brightness in the SDSS $g$ and $r$ bands for each spaxel was obtained from broadband SDSS images created from the data cube.
The measured magnitudes are converted to $B$-band magnitudes and  corrected for Galactic foreground extinction using the recalibrated
$A_V$ values from \citet{Schlafly2011} as reported in the NASA/IPAC Extragalactic Database ({\sc ned})\footnote{The NASA/IPAC
Extragalactic Database ({\sc ned}) is operated by the Jet Propulsion Laboratory, California Institute of Technology, under contract with
the National Aeronautics and Space Administration.  {\tt http://ned.ipac.caltech.edu/}}.
The observed surface-brightness profile within a galaxy was fitted by a broken
exponential profile for the disk and by a general S\'{e}rsic profile for the bulge under the assumption that
the position angle of the major photometric axis and the galaxy inclination are  constant within the disk.
The optical radius of the galaxy $R_{25}$ was estimated using the obtained fit.
It should be noted that the observed surface-brightness profile in many galaxies is well  fitted by a broken exponential profile
with negligible contribution from the S\'{e}rsic component.  This indicates that either there is no appreciable bulge in this galaxy
or that the surface brightness of the bulge is described by the exponential profile. If the latter is true, then
the bulge in this galaxy can be missed.

The distances to the galaxies were adopted from {\sc ned}.  The {\sc ned} distances use flow corrections for Virgo, the
Great Attractor, and Shapley Supercluster infall (adopting a cosmological model with $H_{0} = 73$ km/s/Mpc, $\Omega_{m} = 0.27$,
and $\Omega_{\Lambda} = 0.73$).
We have chosen the spectroscopic $M_{sp}$ masses of the SDSS and BOSS \citep[BOSS stands for the Baryon Oscillation Spectroscopic
Survey in SDSS-III, see][]{Dawson2013}.  The spectroscopic masses are from the table {\sc stellarMassPCAWiscBC03} and were determined using the Wisconsin method
\citep{Chen2012} with the stellar population synthesis models from \citet{Bruzual2003}.

The $\sigma$ distributions of the images for around 3000 MaNGA galaxies from the DR15 release were
constructed. We searched for galaxies for further consideration via visual inspection of those maps.
We required that the spaxels with measured emission lines cover an area large enough so that the gas velocity dispersion
distribution could be classified. The distribution of the gas velocity dispersion in a galaxy with high inclination
(close to the edge-on) in particular cannot be classified.  Peculiar and strongly interacting or merging galaxies were excluded. 

We selected 1146 MaNGA galaxies that fit these criteria.
The circumnuclear regions are of the  H\,{\sc ii} type  in 757 galaxies, of the  INT type in 141 galaxies,
of the LINER type in 202 galaxies, and of the  Seyfert type in 46 galaxies.

\section{Two types of $\sigma$ distribution in galaxies}

\begin{figure}
\resizebox{1.00\hsize}{!}{\includegraphics[angle=000]{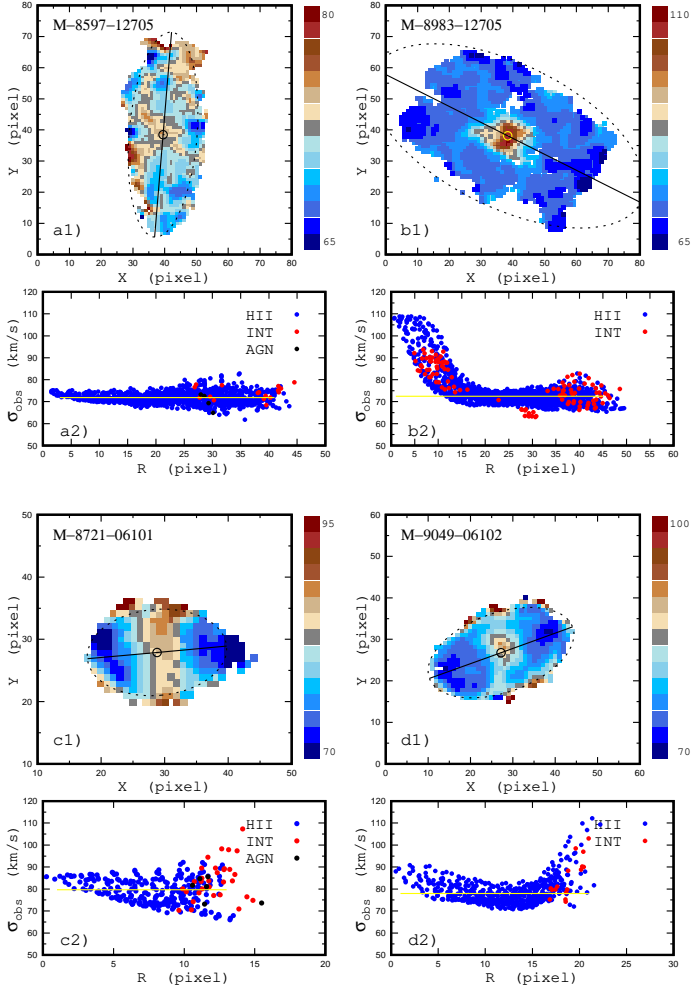}}
\caption{
  Examples of distributions of the gas velocity dispersion $\sigma$ over images of galaxies.
  {\em Panel} $a1$: Galaxy with R distribution of $\sigma$ without an appreciable enhancement at the center.
  The values of the gas velocity dispersion are color-coded.
  The circle shows the photometric center of the galaxy, the solid line indicates the position of the major photometric
  axis of the galaxy, and the dashed line marks the optical radius of the galaxy, $R_{25}$.
  {\em Panel} $a2$: Observed gas velocity dispersion as a function of radius for individual spaxels.
  The BPT type of the spectra is color-coded. The line indicates the median value of the gas velocity dispersions.
  {\em Panel} $b1$: Galaxy with R distribution and with a significant enhancement of the $\sigma$ at the center
  {\em Panel} $b2$: Radial distribution of the $\sigma$ in this galaxy.  
  {\em Panels} $c1$ and $d1$: Examples of galaxies with B distribution. 
  {\em Panels} $c2$ and $d2$: Gas velocity dispersion as a function of radius for those galaxies.
}
\label{figure:map}
\end{figure}

There is a variation in the values of the gas velocity dispersion across the image of each galaxy.
The variation of $\sigma$ can be chaotic or can show a global structure.  
If the amplitude of the chaotic variation of the $\sigma$ is more or less similar over the whole image
of the galaxy, then the distribution of the $\sigma$  in this galaxy can be considered as a uniform distribution
in the first approximation. Panel a1 of Fig.~\ref{figure:map} shows an example of the uniform distribution of $\sigma$
over the image of the galaxy, and panel a2 shows the $\sigma$ as a function of radius for this galaxy.
The value of the gas velocity dispersion is usually enhanced at the center of the galaxy.
Panel b1 of Fig.~\ref{figure:map} shows an example of the distribution of the $\sigma$ with a significant enhancement at the center, 
and panel b2 shows the $\sigma$ as a function of radius.
The global distribution of the $\sigma$ in the galaxy presented in  panel b1 of Fig.~\ref{figure:map} 
shows a radial symmetry. The distribution of the $\sigma$ of such a type (radial symmetry in the  $\sigma$ distribution) 
will be denoted as "R distribution." 
The uniform distribution of the $\sigma$ is a particular case of R distribution. 
By visually inspecting the $\sigma$ maps in our sample of galaxies, we found that the $\sigma$
distribution in 909 of the 1146 galaxies can be classified as R distribution.    

A prominent feature in the distribution of the gas velocity dispersion in some galaxies is a band of
enhanced $\sigma$ along the minor axis of the galaxy. The band can be found both in galaxies with and without
the enhancement of the $\sigma$ at the center. 
Panels c1 and d1 of Fig.~\ref{figure:map} show examples of galaxies with bands of enhanced $\sigma$
in the distribution of the gas velocity dispersion across the image of the galaxy.   Panels c2 and d2 of
Fig.~\ref{figure:map} show the $\sigma$ as a function of radius for those galaxies.
Such a distribution of the gas velocity dispersion in the galaxy (band in the  $\sigma$ distribution)
will be denoted as "B distribution." 
We found that the $\sigma$ distribution in 159 out the 1146 galaxies of our sample can be classified as B distribution.    
It should be emphasized that the band of enhanced $\sigma$ in all the galaxies with B distribution
is close to the minor axis of the image, with one exception. The band of enhanced $\sigma$ in MaNGA
galaxy M-8554-06101 is located along the major axis of the image; this galaxy is not included in the list
of galaxies with B distribution. 

We find it difficult to classify the distribution of the gas velocity dispersion
in 78  of the galaxies from our sample. In some cases, we cannot chose whether the distribution of the  $\sigma$
corresponds with R distribution or B distribution; the visual classification of the distribution of
the gas velocity dispersion across the image of the galaxy is somewhat arbitrary. 
In other cases, there is a feature in the distribution of the gas velocity dispersion that makes the
$\sigma$ distribution peculiar. 

The line-of-sight velocity field serves to examine the galaxy rotation. Accurate rotation curves cannot be determined
for many galaxies from our sample because the curves of isovelocities in the line-of-sight velocity fields in some galaxies
more resemble a set of straight lines than a set of parabola-like curves (the hourglass-like picture for the rotation disk).
Therefore, we restricted ourselves by the qualitative analysis of the galaxy rotation, that is, we carried out a visual
inspection of the line-of-sight map with the aim of establishing whether the galaxy is rotating. 
We found that all the galaxies from our sample are rotating objects. 

For the purpose of comparing the properties of the galaxies with R and B distributions of gas velocity dispersions,
we selected a sample of galaxies with reliably determined parameters using the following four criteria.
First, galaxies with non-classified distribution were excluded.\ Second,  galaxies with
circumnuclear regions of the Seyfert type were also excluded. 
The emission line profiles of the spaxel spectra were fitted by single Gaussians. 
A broad component can be expected in the emission line profiles in the spaxel spectra
of the Seyfert type.  Hence the single Gaussian is not an adequate approximation
for the profile of the emission line with a broad component; such lines should be fitted by two Gaussians.
Third, galaxies with an INT-type circumnuclear region were excluded
because their contribution from the AGN and SF to the radiation of the circumnuclear regions
can significantly change from galaxy to galaxy.
Fourth,  galaxies where the spaxels with available spectra cover less than  two-thirds of the optical radius
were rejected because the determination of parameters (e.g., optical radius $R_{25}$ or the median value of the gas velocity
dispersion $\sigma_{m}$) are questionable.
The selected sample includes 806 galaxies: 687 galaxies with R distribution and 119 galaxies with B distribution.

\begin{figure}
\begin{center}
\resizebox{1.000\hsize}{!}{\includegraphics[angle=000]{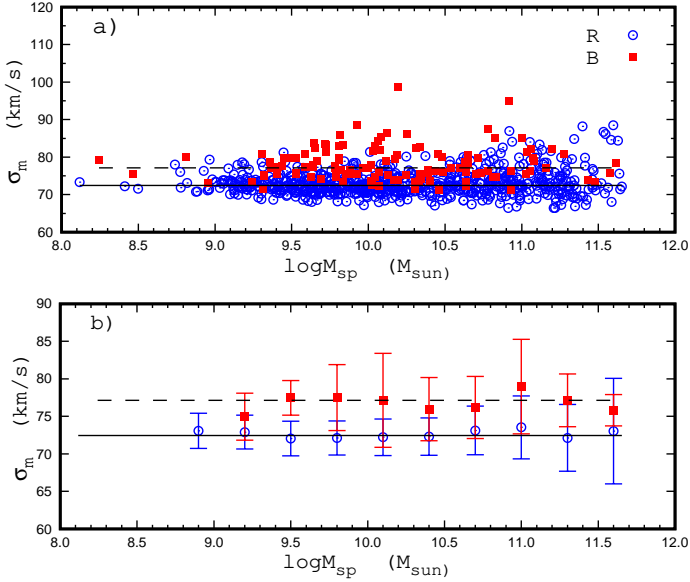}}
\caption{
Median values of gas velocity dispersion $\sigma_{m}$ in galaxies with R  and B distributions. 
{\em Panel} $a$:  $\sigma_{m}$ as a function of the spectroscopic stellar mass $M_{sp}$ for
individual galaxies with
R distribution (blue circles) and with B distribution (red squares).
The solid line shows the median value of $\sigma_{m}$ in galaxies with R distribution, and
the dashed line shows the median value of $\sigma_{m}$ for galaxies with B distribution.
{\em Panel} $b$: Median values of $\sigma_{m}$ for galaxies in
bins of 0.3 dex in $M_{sp}$ for R-distribution galaxies (blue circles) and B-distribution galaxies (red squares).
The lines come from panel $a$.
}
\label{figure:msp-svm}
\end{center}
\end{figure}

\begin{figure}
\begin{center}
\resizebox{1.000\hsize}{!}{\includegraphics[angle=000]{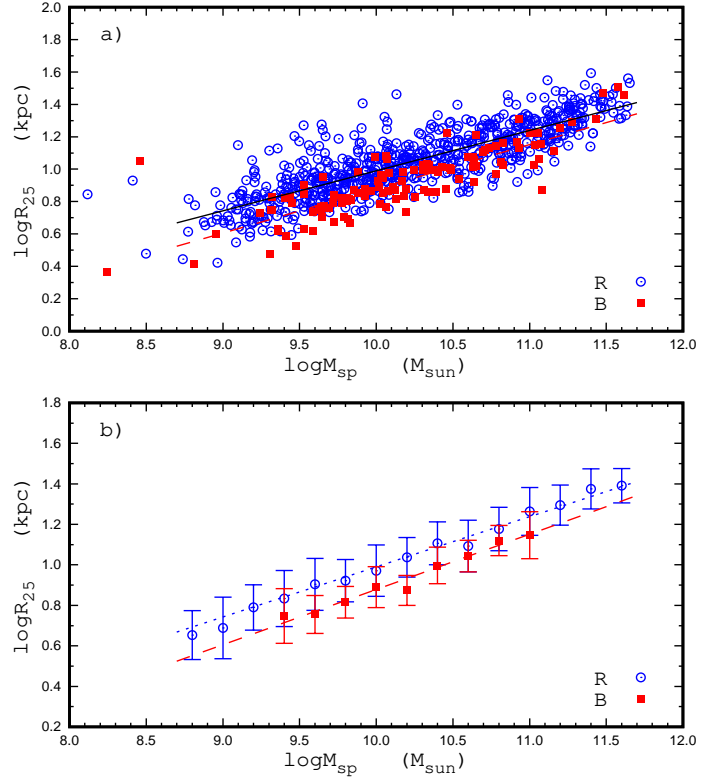}}
\caption{
  Relation between optical radius $R_{25}$ and stellar mass $M_{sp}$.
  {\em Panel} $a$: $R_{25}$ as a function of $M_{sp}$ for individual galaxies with R distribution (blue circles)
  and B distribution (red squares) of the gas velocity dispersion.
  {\em Panel} $b$: Median values of $R_{25}$ for galaxies in
  bins of 0.3 dex in $M_{sp}$ for galaxies with R distribution, shown with blue circles.
  The solid line is the linear best fit to those data. Red squares denote the binned median values for galaxies with
  B distribution. The dashed line is the linear best fit to those data.  
}
\label{figure:msp-ro}
\end{center}
\end{figure}

\begin{figure}
\begin{center}
\resizebox{1.000\hsize}{!}{\includegraphics[angle=000]{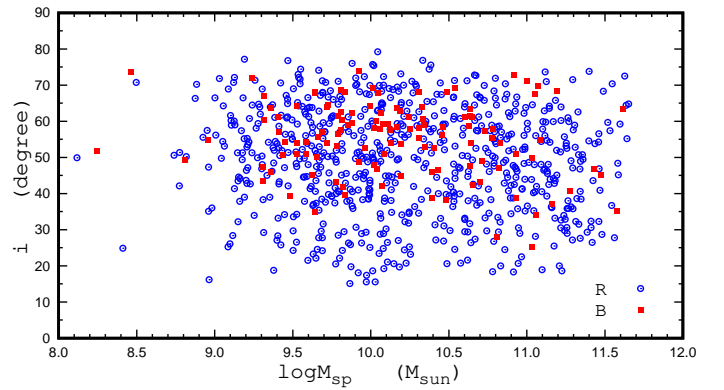}}
\caption{
  Inclination angle $i$ as a function of stellar mass $M_{sp}$ for galaxies with R distribution (blue circles)
  and B distribution (red squares) of the gas velocity dispersion.
}
\label{figure:msp-i}
\end{center}
\end{figure}

Panel a of Fig.~\ref{figure:msp-svm}  shows the median value of the observed gas velocity dispersion $\sigma_{m}$
as a function of spectroscopic stellar mass $M_{sp}$ for galaxies with R distribution (blue circles) and
for galaxies with B distribution (red squares).
Panel b of Fig.~\ref{figure:msp-svm}  represents  the median values of $\sigma_{m}$ for galaxies in
bins of 0.3 dex in $M_{sp}$ for galaxies with R distribution (blue circles) and
for galaxies with B distribution (red squares).
The median value of the $\sigma_{m}$ for the 687 galaxies with R distribution is equal to 72.45 $\pm$ 3.24 km/s
(solid lines in panels a and b of Fig.~\ref{figure:msp-svm}). This value agrees with the median value of the $\sigma_{m}$
 (72.5 $\pm$ 2.8 km/s) obtained for the 161 galaxies in \citet{Pilyugin2020b}.
The median value of the $\sigma_{m}$ for the 119 galaxies with B distribution is equal to 77.15 $\pm$ 4.66 km/s
(dashed lines in panels a and b of Fig.~\ref{figure:msp-svm}).
Thus, the median value of the gas velocity dispersion
in the galaxy is related to the type of the $\sigma$ distribution across the image of the galaxy, in the sense that the median
value of the observed gas velocity dispersion in galaxies with B distribution is higher, on average, by around 5 km/s
compared to galaxies with R distribution.

Next, we examined whether other properties of galaxies are related to the type of  $\sigma$ distributions.  
Blue points in panel a of Fig.~\ref{figure:msp-ro}  indicate the optical radius $R_{25}$ as a function of stellar mass $M_{sp}$
for individual galaxies with R distribution. The linear least-squares fit to those data is
\begin{equation}
\log R_{25} = 0.248(\pm0.006)\log M_{sp} - 1.488(\pm0.065) 
\label{equation:msp-ro-r}
\end{equation}
and is shown  by the solid line in panel $a$ of Fig.~\ref{figure:msp-ro}. The correlation between the optical radius $R_{25}$ and the stellar mass
for galaxies with R distribution is rather tight. The correlation coefficient is 0.83. The mean value of the scatter around
the $R_{25}$ -- $M_{sp}$ relation is 0.117 dex.
 
Red squares in panel a of Fig.~\ref{figure:msp-ro}  indicate the optical radius $R_{25}$ as a function of stellar mass $M_{sp}$
for individual galaxies with B distribution. The linear least-squares fit to those data is
\begin{equation}
\log R_{25} = 0.273(\pm0.016)\log M_{sp} - 1.849(\pm0.164) 
\label{equation:msp-ro-b}
\end{equation}
and is shown  by the dashed line in panel $a$ of Fig.~\ref{figure:msp-ro}. The correlation between the optical radius $R_{25}$ and the stellar mass
for galaxies with B distribution is also tight. The correlation coefficient is 0.84. The mean value of the scatter around
the $R_{25}$ -- $M_{sp}$ relation is 0.110 dex.
 
Panel b of Fig.~\ref{figure:msp-ro} represents the median values of $R_{25}$ for galaxies in
bins of 0.3 dex in $M_{sp}$ for galaxies with R distribution (blue circles) and for galaxies with B distribution (red squares).
The lines are the same as  in panel a of Fig.~\ref{figure:msp-ro}.  
An examination of  Fig.~\ref{figure:msp-ro} shows that the optical radius of the galaxy $R_{25}$ is related to the type of
 $\sigma$ distribution across the image of the galaxy, in the sense that the optical radius in a galaxy with B distribution
is lower, on average, by around 0.1 dex compared to a galaxy of a similar mass  with R distribution.

Thus, the median value of the gas velocity dispersion in a galaxy is related to the type of  $\sigma$ distribution across the image of
the galaxy. The optical radius of a galaxy $R_{25}$ is also related to the type of  $\sigma$ distribution.
This suggests that the presence of the band of  enhanced gas velocity dispersion  is an indicator of a specific evolutionary stage in the galaxy.  

The  band of  enhanced $\sigma$ in a galaxy with B distribution is located along the minor
axis of the galaxy. This can be explained if there is an asymmetry in the gas velocity dispersion,
in other words, if the radial component of the gas velocity dispersion is larger than the azimuthal and vertical
(perpendicular to the galaxy plane) components.

Since the band of enhanced gas velocity dispersion is located along the minor axis of the image of the galaxy,
a question arises of whether or not the band can be an artificial effect
(i.e., if the band can appear due to the high inclination of the galaxy).  Figure~\ref{figure:msp-i} shows 
the inclination angle -- stellar mass diagram for galaxies with R and B distributions. Inspection of Fig.~\ref{figure:msp-i} shows
that there are many galaxies with high inclinations (up to and even in excess of 70$\degr$) where the band of enhanced $\sigma$ is not
revealed. At the same time, the bands are revealed in galaxies with lower inclinations. This indicates that the band is not an 
artificial effect in galaxies with high inclination. 
However,  the band of enhanced $\sigma$ is found more often in galaxies with high inclination. The median value of the inclination of galaxies
with B distribution is 56.6$\degr$, with a scatter of 10.2$\degr$.  The median value of the inclination of galaxies with R distribution
is 50.2($\pm$14.4)$\degr$ for our sample of galaxies.

The mass of  galaxies with AGN is usually higher than 10$^{10.5}$ $M_{\odot}$, while 
the mass of  galaxies with B distribution is often lower than 10$^{10.5}$ $M_{\odot}$.
This suggests that the B-distribution phenomenon is not related to the AGN phenomenon. 
Indeed, the LINER circumnuclear region is revealed only in four out of the 119 galaxies with B distribution,
while the LINER is revealed in 151 out of the 687 galaxies with R distribution.
A bulge is found in several galaxies with B distribution only. 
This indicates that B distribution is not related to the bulge.

\section{Conclusions}

The  parameters of emission lines in spaxel spectra of  MaNGA galaxies are estimated using a Gaussian fit to the line profiles.
The sigma of the best-fit Gaussian to the H$\alpha$ line is converted into the gas velocity dispersion $\sigma$.
The distributions of the gas velocity dispersion $\sigma$ across the images of 1146 MaNGA galaxies
are constructed. We find that there are two types of  gas velocity dispersion distributions
across the images of galaxies. The distributions of 909 of the galaxies show a radial symmetry with
or without the enhancement of the $\sigma$ at the center (R distribution). There is a band of enhanced  $\sigma$ along the
minor axis in the images of 159 of the galaxies with or without the enhancement of the $\sigma$ at the center
(B distribution). The $\sigma$ distribution across the images of 78 of the galaxies cannot be reliable classified.

We selected 806 galaxies with the best defined characteristics 
and compared the properties of  galaxies with R distribution (687) to  galaxies with B distribution (119).
We find that the median value of gas velocity dispersion $\sigma_{m}$ in galaxies with B distribution
is higher, on average, by around 5 km/s than that in galaxies with R distribution. At the same time, the value of $\sigma_{m}$
is similar for galaxies of different masses for a given type of $\sigma$ distribution.

We also find that the optical radius $R_{25}$ of  galaxies with B distribution is lower by around 0.1 dex, on average,
than that of  galaxies of similar masses with R distribution.
Thus the properties of galaxies are related with the type of the  distribution of the gas velocity dispersion $\sigma$ across their image. 
This suggests that the presence of the band of  enhanced gas velocity dispersion is an indicator of
a specific evolution (or specific stage in the evolution) of a galaxy. 

The presence of the band of  enhanced gas velocity dispersion is not related to the AGN phenomenon. 
The band of enhanced  $\sigma$ along the minor axis of the galaxy can be explained if there is an asymmetry in the gas
velocity dispersion, that is, if the radial component of the gas velocity dispersion is larger than the azimuthal and vertical
(perpendicular to the galaxy plane) components.
    
\section*{Acknowledgements}

We are grateful to the referee for his/her constructive comments. \\
We thank Dr. S.G.Kravchuk for useful discussion.  \\ 
L.S.P acknowledges support within the framework of the program of the NAS of
Ukraine “Support for the development of priority fields of scientific
research” (CPCEL 6541230)" \\
I.A.Z acknowledges support by the grant for young scientist’s research
laboratories of the National Academy of Sciences of Ukraine. \\
The work is performed according to the Russian Government Program of Competitive Growth
of Kazan Federal University and Russian Science Foundation, grant no. 20-12-00105 \\
We acknowledge the usage of the HyperLeda database (http://leda.univ-lyon1.fr). \\
Funding for SDSS-III has been provided by the Alfred P. Sloan Foundation,
the Participating Institutions, the National Science Foundation,
and the U.S. Department of Energy Office of Science.
The SDSS-III web site is http://www.sdss3.org/. \\
Funding for the Sloan Digital Sky Survey IV has been provided by the
Alfred P. Sloan Foundation, the U.S. Department of Energy Office of Science,
and the Participating Institutions. SDSS-IV acknowledges
support and resources from the Center for High-Performance Computing at
the University of Utah. The SDSS web site is www.sdss.org. \\

\end{document}